# A W-test collapsing method for rare variant testing with applications to exome sequencing data of hypertensive disorder

Rui Sun[1+], Haoyi Weng[1+], Inchi Hu[2], Junfeng Guo[13], William K.K. Wu[4], Benny Chung-Ying Zee[1*], Maggie Haitian Wang[1*]

1 Division of Biostatistics and Centre for Clinical Research and Biostatistics, JC School of Public Health and Primary Care, the Chinese University of Hong Kong, Shatin, N.T., Hong Kong SAR; the CUHK Shenzhen Research Institute, Shenzhen, China

2 ISOM Department and Biomedical Engineering Division, the Hong Kong University of Science and Technology, Kowloon, Hong Kong SAR

3 the Australian National University, Canberra, Australia

4 Department of Anesthesia and Intensive Care, the Chinese University of Hong Kong, Hong Kong, Hong Kong SAR.

[+] These authors contributed equally to this work

[*] Correspondence: maggiew@cuhk.edu.hk

## Abstract

Advancement in sequencing technology enables the study of association between complex disorders and rare variants with low minor allele frequencies. One of the major challenges in rare variant testing is lack of statistical power of traditional testing methods due to extremely low variances of single nucleotide polymorphisms. In this paper, we introduce a W-test collapsing method that evaluates the distributional differences in cases and controls using a combined log of odds ratio. The proposed method is compared with the Weighted-Sum Statistic and Sequence

Kernel Association Test using simulation data sets and showed better performances and faster computing speed. In the study of real next generation sequencing data set of hypertensive disorder, we identified genes of interesting biological functions that are associated to metabolism disorder and inflammation, which include the *MACROD1*, *NLRP7*, *AGK*, *PAK6* and *APBB1*. The W-test collapsing method offers a fast, effective and alternative way for rare variants association analysis.

**Background**

In the past decade, genetic association studies identified a repertoire of associated variants for complex disorders [1]. However, a large portion of the disease heritability remains unexplained. Areas that may account for the missing heritability in complex traits include polygenetic effects, gene-environment interactions, and rare variants associations [2]. Sequencing technology development in recent years allow deep DNA sequencing to be done at lower cost, and the Next Generation Sequencing (NGS) is made available for studying extreme low frequency variants. One main challenge in studying rare variants association is the lack of statistical power due to low minor allele frequency (MAF). Furthermore, the high volume of genetic markers also increases the burden of multiple testing. A number of statistical methods for rare variant association testing have been proposed, which can be generally divided into two categories, the burden tests such as the Weighted Sum Statistic (WSS) [3] and Combined Multivariate Collapsing [4]; and the variance component tests such as the C-alpha test [5] and the Sequence Kernel Association Test (SKAT) [6]. Both types of methods increase power for rare variants by defining larger genomic regions to conduct the test. In this paper, we introduce a W-test collapsing method to evaluate rare variant data. The method tests the distributional differences in cases and controls using a retrospective design after collapsing the allele frequencies within a genomic region。 The test statistic carries an inherent Chi-squared distribution with data-set dependent degrees of freedom. Power and type I error rate of W-test collapsing is compared with the WSS and the SKAT using simulated data sets. The proposed method showed better performances and several hundred times faster in computing speed. The proposed method is also applied on real NGS data set of hypertensive disorder, and interesting genes have been found.

**Method**

*The W-test*

The W-test is formulated to test the distributional difference of a SNP in the affected from the unaffected group [7]. Suppose a variant X has $k$ levels, and Y is binary. If a variant carries three genotypes, *AA*, *Aa* and *aa*, then $k=3$. The test takes the following form:

$$W = h \sum_{i=1}^{k} \left[ \log \frac{\hat{p}_{1i}/(1-\hat{p}_{1i})}{\hat{p}_{0i}/(1-\hat{p}_{0i})} \bigg/ SE_i \right]^2 \sim \chi_f^2 \qquad \text{Equation 1}$$

$$SE_i = \sqrt{\frac{1}{n_{0i}} + \frac{1}{n_{1i}} + \frac{1}{N_0 - n_{0i}} + \frac{1}{N_1 - n_{1i}}},$$

where $\hat{p}_{1i}$ is the proportion of cases in cell-$i$ out of total case number, and $\hat{p}_{1i}$ is the proportion of controls in cell-$i$ out of total control number. $SE_i$ is the standard error of log odds ratio of cell-i, in which $n_{1i}$ and $n_{0i}$ are the number of cases and controls in the $i^{th}$ cell; $N_1$ and $N_0$ are the total number of cases and controls, respectively. The scalar $h$ and the degrees of freedom parameter $f$ are obtained by estimating the covariance matrix from the bootstrapped data under null hypothesis. The statistic follows a Chi-squared distribution with $f$ degrees of freedom. Empirical studies give $h \approx (k-1)/k$ and $f \approx k-1$. When sample size and the number of markers are both around 1000, the estimated parameters begin to converge at bootstrap times greater than 200 [7]. The *h* and *f* of simulation and real data set can be found in Supplementary Materials S1. The W-test is especially powerful under low frequency variable environment (when MAF is between 1-5%) [7], as the data-dependent parameters help to reduce bias in test probability distribution arise from small MAFs.

*The W-test collapsing method*

The W-test collapsing method is a direct extension of the original W-test on rare variants. Suppose a genomic region contains *m* rare SNPs; each SNP can form a contingency table. The *m* contingency tables of the SNPs in the genomic region are summed cell by cell, and a combined contingency table is formed for this collapsing region. The W-test collapsing applies the original W-test on top of the combined contingency table as a new statistic, which follows a Chi-squared distribution with *f* degrees of freedom. The *h* and *f* are estimated from the data under the null hypothesis based on the collapsed region.

*Comparison with other rare variant methods*

Two representative rare variant methods are considered as alternative approaches, namely, the SKAT and the WSS [3, 6]. The SKAT is a kernel machine regression method that incorporates a variance component score for coefficient evaluation. It has the advantage of dealing with both continuous and discrete phenotypes, and can test genetic effect in opposite directions [8]. The WSS first gives each individual a weighted sum score of mutations counts, and then test for excess of mutations in cases compared to null hypothesis [9]. Permutations are needed to calculate p-values for the WSS.

*Simulation data*

Simulation data are used to evaluate power and type I error rate of different methods. Each replicate includes 1,920 rare SNPs and 2,000 subjects. Rare variants are randomly generated to have MAF range between 0.01% and 1%. One gene is consisted of 32 SNPs. Each replication data includes 60 genes, in which 10 genes contain causal SNPs. The phenotypes are generated by a logistic regression model containing all the causal SNPs and a random error term [10]. Two phenotypic models are considered:

*Scenario I*: In a causal gene, 12 causal SNPs cluster together in the same effect direction;

*Scenario II*: In a causal gene, 8 causal SNPs cluster together in opposite effect directions, with 6 SNPs of risk effect and 2 SNPs of protective effect.

There are 37.5% of causal variants in Scenario I, and 25% in Scenario II. It is known that Scenario I model favors burden-like test, and scenario II is suitable to apply the variance component test [11]. Five hundred replicated data sets are generated to calculate the power and type I error rate. Power is the averaged true positive rates in 500 replicates, and type I error rate is the average false positive rates. A positive gene is defined as the gene which p-value is smaller than Bonferroni corrected alpha of 5% in 60 genes.

*GAW18 simulated and real data sets*

Both simulated and real data set from the Genetic Analysis Workshop 18 (GAW18) are applied. The subjects are unrelated Mexican Americans who are enriched in type 2 diabetes, drawn from the T2D-GENES consortium project 2 [10]. The GAW18 simulated data sets contain real genotype data and predefined systolic and diastolic blood pressure (SBP and DBP) generated from a linear regression model [10]. Hypertension is defined at SBP>90 mm Hg or DBP>140 mm Hg. We use rare variants (MAF<1%) on chromosome 3 and 200 replicates for power and type I error rate calculation. The simulated data sets consist of 330 cases, 1600 controls and 42,825 rare SNPs. The total number of causal SNPs is 164. The rare variant methods need to be applied based on a certain genomic region, while the optimal regions are non-identical for different methods. We estimate optimal collapsing window for each method at which they have the best power for the GAW18 simulated dataset. The optimal window for the SKAT and the W-test is 15 SNPs, and for WSS is 10 SNPs. A causal region is defined as the genomic area containing at least one causal SNP. Receiver operating characteristic (ROC) curve is plotted using the top number of collapsed regions. For real data analysis, there are 398 hypertensive individuals and 1,453 controls. Quality control (QC) is conducted to remove variants with missing value percentage over 5% and inconsistent genotyping format. Odd numbered chromosomes are evaluated and the total number of rare SNPs passed QC is 308,722 in the real data set. The collapsing window size for real data is 15, so the number of multiple tests is 308,722/15=25,385, and Bonferroni corrected significance level at 5% alpha is $1.97 \times 10^{-6}$.

**Results**

*Comparison of alternative methods in simulation data*

In Scenario I where causal SNPs in a gene have the same effect direction, the W-test's power is 66.6%, slightly better than WSS's 66.3%. Both burden tests outperform the SKAT's power 55.9% (Table 1). For Scenario II, where the causal markers show different effect directions, the SKAT's power is the highest, 93.0%, followed by W-test's 47.1%, and WSS's 39.6%. All methods' type I error rates are conservative: 0.13% for W-test, 0.13% for SKAT, and 0.17% for WSS (Table 1). The W outperforms WSS in both scenarios. Furthermore, the W-test takes 0.06 seconds to evaluate 10 genes; it is 235 times faster than SKAT, and 393 times faster than the WSS. The W-test benefits from its intrinsic probability distribution estimated from the small bootstrapped samples to calculate p-values, compared to other rare variant association tests that require complete permutation or Monte Carlo estimation.

*Application to GAW18 simulation study*

The ROC curves of the W-test, SKAT and the WSS are plotted in Figure 1. The figure shows that in the GAW18 data set, all methods have low power and high false positive rates. Similar lack of power has been observed by other studies on the same data [11-13]. Nevertheless, the W-test performs the best among the three methods. At false positive rate 52.5%, the W-test collapsing has true positive rate (TPR) 57.3%, which is 52% for the SKAT and 52.8% for the WSS. The causal SNPs distribution in the top ranked causal genes is exhibited in Table 2. All methods are able to find extreme rare variants with MAF 0.0003 and SNPs of very small effect sizes. The characteristics of identified causal markers are also intriguing: Except for one gene *ZBTB38* that is identified by all three methods, the regions short-listed by SKAT and WSS share no similarities, while the W-test found common regions to the other two methods (Table 2). The causal regions identified by SKAT are mostly composed of a single SNP with very large effect sizes (coefficients with absolute value ranges from 0.06 to 20); and the WSS identified regions containing two or more causal SNPs of moderate effect size (coefficients in the linear regression with absolute magnitude under 1.5). Interestingly, the W-test collapsing identified both the moderate effect sizes genes *SEMA3F* and *MUC13*, and the unique genes *SENP5* of a large effect size. These results showed that the W-test collapsing has slight power advantage than the SKAT and WSS in the GAW18 data; it also shares common properties of the other two distinct methods, and has unique finding of its own.

*Application to real hypertension exome sequencing data*

We applied the W-test collpasing method on real exome data of hypertensive disorder. One region reached bonferroni corrected significance level with p value $6.1\times10^{-7}$. Not surprisingly, SNPs in this region are not discoverable by single marker test as they are individually non-significant. The identified chromosome position contains the gene *MACROD1/LRP16* (11q11, average MAF = 0.001, odds ratio for collapsed marker = 3.84), which is a ubiquitous protein module that binds ADP-ribose derivatives, and supports many different protein functions and pathways. It was reported that the *LRP16* is over expressed in tissues of colorectal and gastric carcinoma patients [14, 15]. The top 17 regions that have large to moderate effect are listed in Table 3. These include genes *NLRP7*, *AGK*, *PAK6* and *APBB1*, which have potential association to hypertension. The *NLRP7* (MAF = 0.0027, OR=2.23, W-test collapsing p-value = $8.3\times10^{-6}$) encodes a protein that is implicated in the activation of proinflammatory caspases through multiprotein complexes inflammasones. Studies reported that this gene is associated with molar pregnancy and other pregnancy complications [16, 17]. The acylglycero kinase (*AGK*) is involved in lipid and glycerolipid metabolism, and it was found to have a significant over expression in retinas of diabetic rats [18], and may play a role in the development of cataract [19]. The gene *PAK6* is a p21-activated kinase that is central to signal transduction and cellular regulation. Previous cell-line, tissue and gene expression studies reported this gene may play essential roles in the initiation and progression of hepatocellular carcinoma [20, 21]. The protein encoded by *APBB1* is a member of the Fe65 protein family, and interacts with the transcription factor *LBP1* and the low-density lipoprotein receptor-related protein [22].

**Discussion**

We propose a W-test collapsing method to test the association between a dichotomous phenotype and rare genetic variants. It is model-free, fast, and tests the distributional differences in cases and controls through the integrated log odds ratios. Because of the odds ratio design, it has a unique retrospective design that is suitable to be applied on both prospective and retrospective data sets. The proposed method can be categorized as a burden test; therefore, it is more advantages under the scenario when the SNPs effect directions are the same, compared to variance component test. It outperforms another burden test WSS under different effect scenarios. The advantage of the proposed method, apart from power and controlled type I error rate, is its p-value calculation free from large permutation. There are two major beneficial aspects: first is the extraordinary computing speed. Almost all rare variant tests heavily rely on Monte Carlo or permutations to calculate p-values. And the computing burden prohibits possible optimization of collapsing region in whole exome sequencing data. Second, the test inherits a dataset-dependent probability distribution. The proposed method uses small bootstrapped samples under null hypothesis to estimate refined degrees of freedom (non-integer) for the testing probability distribution. Because the estimation involves data covariance structure, the resulting chi-squared distribution can corrects for potential bias due to complex data structures and can give accurate p-value calculation at minimum computing cost.

We compared different methods at their optimized collapsing region on the GAW18 data set, which is usually not performed in the literature. The optimal bin size is related to the number of causal markers in the region, causal marker's effect sizes and directions, weighting scheme and the distribution of mutations in cases and controls. This study demonstrated that the optimal window sizes are not the same for different methods; and the best collapsing bin is smaller than the commonly adopted range, such as a gene or pathway [3, 6]. Further study is needed to explore how to locate the best collapsing region in real exome sequencing data.

In terms of identified causal regions in GAW18 data, the W-test collapsing method shared similarities with the WSS that they both identified the regions populated with many causal variants. However, there are differences in the two burden methods: the WSS mainly found variants with moderate effect size, but the W-test collapsing identified large negative effect variants as well. The reason can be explained from the formulation of the two statistics. The WSS adds the number of mutated alleles in a genomic region and weight them inversely by the proportion of mutations in the unaffected subjects. A critical assumption of the WSS is that the minor alleles are mutations and contribute to disease risk [3]; therefore, if the minor allele with protective effect are concentrated in the unaffected, the WSS will down weight the variants and could miss them. As a result, the causal gene *SENP5* that contains 5 large negative effects SNPs is not short-listed by the WSS in the GAW18 simulation study, in which most of the mutations in this gene occur in the unaffected only once (Table 2). On the contrary, the W-test collapsing does not make assumption on mutation effect; it directly tests the distributional differences between the affected and unaffected; therefore, it is more general than the WSS to identify protective effect rare variants. The SKAT performs a kernel regression using variance component in each region, and variants with small MAF will be given heavier weights. The SKAT method tends to select a region that includes a few rare variants with large effect sizes, and it allows variants within a region to adopt opposite effect directions. In the GAW18 simulation study, four out of the five top regions identified by SKAT contain only 1 causal marker, and the rest gene *ARHGEF3* contains 2 SNPs of same effect direction. Except for the first gene, the SKAT does not share common identified regions to the WSS and the W-test collapsing method. The rest variation in the causal marker distribution relates to the noise SNPs contained in collapsing region, and the random variations in the replication data sets. The simulation study showed that the proposed method has advantage in locating clustered causal rare variants, and is more general than the WSS to detect protective variants. Nevertheless, the three tests discussed in this study all have distinct properties and strength. In real data analysis, when the underlying genetic model is unknown, the methods may need to be considered jointly to obtain a complete picture. There are existing combined test such as the SKAT-O and Fisher's method, which pool the individual tests by some tuning parameters [23, 24]. These tests are more appropriate when the underlying model is unknown, but less powerful for specific scenarios [25]. The W-test collapsing method can be naturally pooled with the SKAT by the Fisher's method, which we might explore in future study.

To conclude, we proposed a W-test collapsing method to evaluate rare variant in exome sequencing data, thus enables the testing of the whole genome data under an integrated statistical framework. The proposed method offers a very efficient and effective way for rare variant association analysis.

**Acknowledgements**

This work has been supported the Chinese University of Hong Kong Direct Grant [4054169] to MHW; Research Grant Council – General Research Fund [476013] to MHW; and National Science Foundation of China [81473035, 31401124] to MHW. This research was conducted using the resources of the High Performance Cluster in Li Ka Shing Institute of Health Sciences, the Chinese University of Hong Kong; and the High Performance Cluster Computing Centre, Hong Kong Baptist University, which receives funding from Research Grant Council, University Grant Committee of the HKSAR and Hong Kong Baptist University. We would like to thank Tony Liu, Sammy Tang, and Morris Law for their technical help in the cluster usage; and Genetic Analysis Workshop 18 for providing the data sets.

**Competing interest**

The authors declare no competing interest.

**Authors' contributions**

MHW conceived the study and wrote the paper, RS and HW processed the data and performed the analysis; JG wrote part of the program; IH and WKKW commented and proved the paper; BZ contributed and coordinated the study.

**Figure Legends**

**Figure 1. ROC of Alternative Methods in GAW18 Simulation Data**

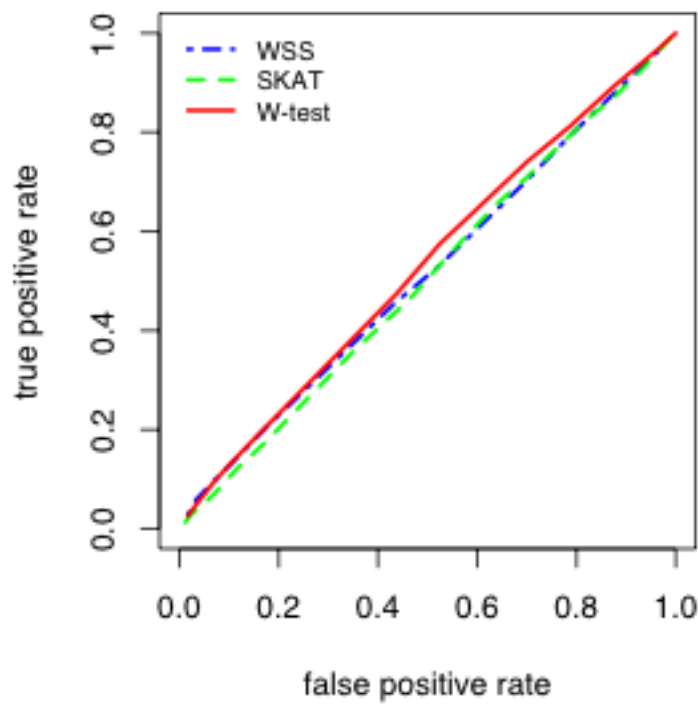

**Tables**

**Table 1. Power and type I error rates of rare variant association tests**

| Statistical tests | Power Scenario I[1] | Power Scenario II[2] | Type I error rate | Speed[3] (seconds) |
|---|---|---|---|---|
| WSS | 66.3% | 39.6% | 0.17% | 23.59 |
| SKAT | 55.9% | 93.0% | 0.13% | 14.13 |
| W-test | 66.6% | 47.1% | 0.13% | 0.06 |

1 Scenario I: causal SNPs clustered together with the same effect direction
2 Scenario II: causal SNPs clustered together, with opposite effect directions
3 Speed is the averaged elapsed time of evaluating 10 genes

**Table 2. Characteristics of identified markers by alternative methods in simulated dataset**

| Method | Rank | Gene | Number of Causal SNPs | Causal SNPs in the gene | | | |
|---|---|---|---|---|---|---|---|
| | | | | position | MAF | SBP effect | DBP effect |
| SKAT | 1 | ZBTB38 | 1 | 141164276 | 0.0003 | -0.007 | -0.002 |
| | 2 | ARHGEF3 | 2 | 56835799 | 0.0003 | -0.067 | -0.062 |
| | | | | 56835795 | 0.0008 | -0.059 | -0.055 |
| | 3 | MAP4 | 1 | 48040284 | 0.0003 | -20.808 | -9.682 |
| | 4 | FLNB | 1 | 58134409 | 0.0005 | 1.687 | 0.249 |
| | 5 | MUC13 | 1 | 124646631 | 0.0003 | 0 | -2.178 |
| WSS | 1 | ZBTB38 | 1 | 141164276 | 0.0003 | -0.007 | -0.002 |
| | 2 | SEMA3F | 3 | 50222143 | 0.0008 | 0.706 | 0.505 |
| | | | | 50214207 | 0.0005 | 0.00002 | 0.00001 |
| | | | | 50222178 | 0.0010 | 0.00007 | 0.00005 |
| | 3 | SEMA3F | 3 | 50222879 | 0.0010 | 1.361 | 0.973 |
| | | | | 50223334 | 0.0003 | 1.010 | 0.722 |
| | | | | 50223764 | 0.0010 | 1.101 | 0.787 |
| | 4 | MLH1 | 2 | 37048495 | 0.0005 | 0 | -0.454 |
| | | | | 37045960 | 0.0008 | 0 | -0.280 |
| | 5 | MLH1 | 2 | 37061893 | 0.0003 | 0 | -0.00004 |
| | | | | 37061929 | 0.0005 | 0 | -0.00001 |
| W-test | 1 | ZBTB38 | 1 | 141164276 | 0.0003 | -0.007 | -0.002 |
| | 2 | SEMA3F | 3 | 50222879 | 0.0010 | 1.361 | 0.973 |
| | | | | 50223334 | 0.0003 | 1.010 | 0.722 |
| | | | | 50223764 | 0.0010 | 1.101 | 0.787 |
| | 3 | MUC13 | 2 | 124632448 | 0.0005 | 0 | -1.244 |
| | | | | 124639097 | 0.0008 | 0 | -0.476 |
| | 4 | SENP5 | 5 | 196612750 | 0.0003 | -4.336 | 0 |
| | | | | 196612959 | 0.0062 | -3.169 | 0 |
| | | | | 196613022 | 0.0003 | -1.697 | 0 |
| | | | | 196613096 | 0.0003 | -4.271 | 0 |
| | | | | 196613191 | 0.0008 | -0.635 | 0 |
| | 5 | SEMA3F | 4 | 50225153 | 0.0003 | 1.418 | 1.013 |
| | | | | 50225255 | 0.0003 | 0.00003 | 0.00002 |
| | | | | 50225285 | 0.0003 | 1.391 | 0.994 |
| | | | | 50225454 | 0.0003 | 0.254 | 0.182 |

**Table 3. Top associated regions in real exome data of hypertensive disorder**

| Rank | Position | Gene | Chr | MAF[1] | OR[1] | W-test p-value |
|------|----------|------|-----|--------|-------|----------------|
| 1 | 64122856-64127883 | *MACROD1/LRP16* | 11 | 0.0010 | 3.84 | $6.1\times10^{-7}$ |
| 2 | 149520895-149521591 | -- | 7 | 0.0012 | 3.09 | $5.2\times10^{-6}$ |
| 3 | 54947030-54947395 | *NLRP7* | 19 | 0.0027 | 2.23 | $8.3\times10^{-6}$ |
| 4 | 5809248-5809272 | -- | 11 | 0.0007 | 1.92 | $9.3\times10^{-6}$ |
| 5 | 40843119-40843343 | -- | 17 | 0.0004 | 8.64 | $9.9\times10^{-6}$ |
| 6 | 115091763-115091808 | -- | 13 | 0.0035 | 0.31 | $1.1\times10^{-5}$ |
| 7 | 91732176-91746400 | -- | 7 | 0.0006 | 4.44 | $2.0\times10^{-5}$ |
| 8 | 67103877-67109739 | *HELZ* | 17 | 0.0003 | 14.8 | $3.0\times10^{-5}$ |
| 9 | 141619479-141635585 | *AGK* | 7 | 0.0015 | 2.55 | $3.6\times10^{-5}$ |
| 10 | 141618669-141619479 | *AGK* | 7 | 0.0012 | 2.80 | $3.9\times10^{-5}$ |
| 11 | 6328839-6413690 | -- | 9 | 0.0017 | 2.22 | $4.1\times10^{-5}$ |
| 12 | 40268764-40290931 | *PAK6* | 15 | 0.0022 | 2.14 | $5.0\times10^{-5}$ |
| 13 | 58102391-58111521 | *ZSCAN18* | 19 | 0.0006 | 4.20 | $5.2\times10^{-5}$ |
| 14 | 41113238-41131622 | *KRTAP9-7* | 17 | 0.0014 | 2.52 | $5.5\times10^{-5}$ |
| 15 | 7794026-7797451 | -- | 17 | 0.0010 | 2.63 | $6.6\times10^{-5}$ |
| 16 | 2435288-2435470 | -- | 11 | 0.0011 | 2.75 | $9.3\times10^{-5}$ |
| 17 | 6341489-6411783 | *APBB1;SMPD1* | 11 | 0.0007 | 3.00 | $9.9\times10^{-5}$ |

1 MAF: average minor allele frequency in the collapsing region; OR: odds ratio for minor allele of the collapsed marker.

# Supplementary Materials

**S1. Estimated *h* and *f***

|  | $k=2$ | $k=3$ |
|---|---|---|
| Simulation data 1 | NA | $h=0.64; f=1.86$ |
| GAW18 simulation data | $h=0.63; f=1.26$ | $h=0.74; f=1.74$ |
| GAW18 real data | $h=0.43; f=0.88$ | $h=0.49; f=1.28$ |